\documentclass[a4paper,11pt]{amsart}
\begin{document}

\title{{\bf  ON DIRAC'S MAGNETIC MONOPOLE}}
\author{Angelo Loinger}
\date{}
\address{Dipartimento di Fisica, Universit\`a di Milano, Via
Celoria, 16 - 20133 Milano (Italy)}
\email{angelo.loinger@mi.infn.it}
%%\thanks{}

\begin{abstract}
The electric charge of the quantization condition of Dirac's
monopole may have any value, we are not obliged to identify it
with the electron charge. Consequently the magnetic charge of the
monopole is quite arbitrary: Dirac's monopole is a mere object of
science fiction.
\end{abstract}

\maketitle

\vskip1.20cm
%\section{}
Too many papers have been written on Dirac's monopole. In recent
years, Dirac's treatment has been reformulated with the language
of differential geometry. However, the fundamental papers remain
those of Dirac \cite{1} and Fierz \cite{2}. In 1987 I have
investigated, with a quick, but conceptually ri\-go\-rous method,
the following significant instance: the motion of a charged
particle in the external magnetic field of a fixed monopole
\cite{3}. This treatment allows to derive the celebrated Dirac's
quantization condition \emph{avoiding Dirac's arbitrary assumption
of the single-valuedness of $\psi$--function}. Thus, my approach
-- as that of Fierz \cite{2} -- is in accord with a basic Pauli's
criterion \cite{4}; moreover, it makes particularly evident that
the electric charge of the particle may have \emph{any} value.
\par
 Let us consider the motion of a particle, having a given mass
 $M$ and an electric charge $Q$, in the magnetic field
 $\bf{B}(\bf{r})=\emph{g } \bf{r}\textrm{/}\mid \bf{r}\mid^{\textrm{3}}$
 generated by a fixed monopole, whose magnetic charge is $g$. (We neglect
 the action on the particle of its own e.m. field.) If
 $\textbf{x}$ is the position vector of the particle with respect
 to an origin coinciding with the monopole, and
 $\bf{\Pi}=\emph{M}\bf{\dot{x}}$ is its kinetic momentum, the
 Hamiltonian $H$ of the problem will be:

\begin{equation} \label{eq:one}
    H=\frac{1}{2M}\bf{\Pi}^{\textrm{2}}.
\end{equation}

We assume the following commutation rules (at equal times):

\begin{equation} \label{eq:two}
    [x_{j},x_{k}]_{-}=0,\:\:\:\:\:(j,k=1,2,3),
\end{equation}

\begin{equation} \label{eq:three}
    [x_{j},\Pi_{k}]_{-}=i\hbar\delta_{jk},
\end{equation}

\begin{equation} \label{eq:eq:four}
    \bf{\Pi}\times\bf{\Pi}=\emph{i} \hbar \frac{\emph{Q}}{\emph{c}}\bf{B}(\bf{x}).
\end{equation}

It is easy to prove that the operator

\begin{equation} \label{eq:five}
    \bf{R}:=\bf{x}\times\bf{\Pi}-\hbar\eta\bf{x}|\bf{x}|^{\textrm{-1}},
\end{equation}

where $\eta$ is the pure number $Qg\textrm{/}(\hbar c)$, commutes
with $H$ -- i.e. is an integral of the motion --, and that

\begin{equation} \label{eq:six}
    \bf{R}\cdot\bf{x}|\bf{x}|^{\textrm{-1}}=\bf{x}|\bf{x}|^{\textrm{-1}}\cdot\bf{R}=-\hbar\eta,
\end{equation}

\begin{equation} \label{eq:seven}
    \bf{R}\times\bf{R}=\emph{i}\hbar\bf{R}.
\end{equation}

The equations of motion of Heisenberg picture are:

\begin{equation} \label{eq:eight}
    M\dot{\bf{x}}=\Pi,
\end{equation}

\begin{equation} \label{eq:nine}
\dot{\bf{\Pi}}=\frac{\emph{Q}}{\textrm{2}\emph{c}}\left(\bf{\Pi}\times\bf{B}-\bf{B}\times\Pi\right).
\end{equation}

By splitting $\bf{\Pi}$ into a radial and a transverse part,
$\bf{\Pi}_{\emph{r}}$ and $\bf{\Pi}_{\tau}$ respectively, we
obtain:

\begin{equation} \label{eq:ten}
    H=\frac{1}{2M}\left(\bf{\Pi}_{\emph{r}}^{\textrm{2}}+\bf{\Pi}_{\tau}^{\textrm{2}}\right),
\end{equation}

\begin{equation} \label{eq:eleven}
    \bf{R}=\bf{x}\times\bf{\Pi}_{\tau}-\hbar\eta\bf{x}|\bf{x}|^{\textrm{-1}},
\end{equation}

\begin{equation} \label{eq:twelve}
    \bf{\Pi}_{\tau}^{\textrm{2}}=\frac{\bf{R}^{\textrm{2}}-\hbar^{\textrm{2}}\eta^{\textrm{2}}}{|\bf{x}|^{\textrm{2}}}.
\end{equation}

Then, by substituting eq.(\ref{eq:twelve}) into eq.(\ref{eq:ten}),
we have:

\begin{equation} \label{eq:thirteen}
    H=\frac{1}{2M}\left(\bf{\Pi}_{\emph{r}}^{\textrm{2}}+\hbar^{\textrm{2}}\frac{\bf{d}^{\textrm{2}}-\eta^{\textrm{2}}}{|\bf{x}|^{\textrm{2}}}\right),
\end{equation}

where $\textbf{d}:=\hbar^{-1}\textbf{R}$ -- and therefore
$\textbf{d}\times\textbf{d}=i\textbf{d}$. To obtain the
quantization condition, one must solve the following eigenvalue
problem:

\begin{equation} \label{eq:fourteen}
    \bf{d}^{\textrm{2}}\chi=\lambda\chi;
\end{equation}

now, it is well known from the representation theory of
three-dimensional rotation group \cite{5} that the most general
eigensolutions of eq.(\ref{eq:fourteen}) dependent only on
\emph{two} parameters (e.g., the polar angles $\vartheta$ ,
$\varphi$) are:

\begin{equation} \label{eq:fifteen}
\left\{ \begin{array}{l} \lambda=j(j+1), \textrm{  with }
j=0,\frac{1}{2},1,\frac{3}{2},\ldots,\\ \\
\chi(\vartheta,\varphi)=\textrm{exp}[i(m_{1}+m_{2})\varphi]P_{m_{1},m_{2}}^{(j)}\cos(\vartheta),
\end{array}\right.
\end{equation}

where $m_{1}$ and $m_{2}$ take the values
$-j$,$-j+1$,\ldots,$j-1$,$j$, and the $P_{m_{1},m_{2}}^{(j)}$'s
are Jacobi polynomials. The quantum number $m_{2}$ gives the
possible values of the orthogonal projection of $\textbf{d}$ onto
$\textbf{x}$; accordingly, we have, owing to eq.(\ref{eq:six}):

\begin{equation} \label{eq:sixteen}
    m_{2}=\eta,
\end{equation}

\begin{equation} \label{eq:seventeen}
    j\geq\eta.
\end{equation}

Eq.(\ref{eq:sixteen}) tells us that

\begin{equation} \label{eq:sixteenbis}
    g=\frac{\hbar c}{2Q}\cdot\textrm{integer},
\end{equation}

Contrary to a diffuse belief, eq.(\ref{eq:sixteenbis}) \emph{has a
poor physical significance}. In fact, since the electric charge
$Q$ may have \emph{any} value, the magnetic charge $g$ of the
monopole is quite arbitrary, in particular it may be \emph{very
small}. Dirac identified $Q$ with the electron charge $e$, but
this is an \emph{unjustified} assumption. Furthermore, as it was
pointed out in 1933 by Bohr and Rosenfeld, macroscopic bodies
(with \emph{large} charges and masses) are essential for the
physical interpretation of quantum electrodynamics.
\par
Solutions (\ref{eq:fifteen}), (\ref{eq:sixteen}) and
(\ref{eq:seventeen}) are formally identical with the solutions of
the quantal problem of an ``infinitely thin'' top, having only
\emph{two}, and not three, degrees of freedom. Such a top
``rotates'' around its axis $\textbf{x}$ with a spin $\hbar\eta$,
while $\textbf{x}$ precesses around the fixed vector $\textbf{R}$.
\par
The above treatment, which is \emph{manifestly gauge invariant},
demonstrates that the physically correct relation between $g$ and
$Q$ is given by eq.(\ref{eq:sixteenbis}). Other quantization
conditions can be found in the literature: they are fully
senseless, because they \emph{depend} both on the mentioned
single-valuedness hypothesis of $\psi$--function and on the number
of the singularity lines extending outward from the particles.

\end{document}